\documentstyle[11pt,newpasp,twoside,epsfig]{article}
\def\edcomment#1{\iffalse\marginpar{\raggedright\sl#1\/}\else\relax\fi}
\reversemarginpar
\begin{document}
\title{PN -- ISM interaction: The observational evidence}
 \author{Florian Kerber}
\affil{ST-ECF, ESO, Garching, Germany}
\author{Elise Furlan}
\affil{Institut f\"ur Astrophysik, Univ. Innsbruck, Austria}
\author{Thomas Rauch}
\affil{Institut f\"ur Astronomie und Astrophysik, Univ. T\"ubingen, Germany}
\author{Miguel Roth}
\affil{Carnegie Observatories, Pasadena, USA}

\begin{abstract}
We present results of an ongoing survey of low surface brightness planetary
nebulae (PNe). Using both narrow-band imaging and long slit spectroscopy we 
have
studied 15 new examples for interaction with the interstellar medium (ISM)
demonstrating that this process is common in evolved PNe.
Characteristic properties of the nebulae in terms of morphology and plasma
parameters have been established and different degrees of interaction
can be identified. Although the study of the objects is at an early stage
the observational findings are in agreement with 
theory. The data indicate that the objects have a very high 
degree of individuality due to the action of complex physics under varying
conditions.
\end{abstract}

\section{Introduction}
\noindent Many aspects of the physics and shapes of
PNe can successfully be
explained in terms of the two-wind 
model by Kwok, Purton, \& Fitzgerald (1978) as the product of the mass-loss 
history of the star on the asymptotic giant 
 branch (AGB) and during the central star evolution.\newline
Our observations indicate that at some point in the evolution of PNe 
other factors may become very important for the further development, and 
for such objects the two-wind paradigm may break down. Old PNe interacting 
with the ISM are a case in point.
This work has implications for a hitherto neglected aspect of PNe: 
they give evidence for an important
process that remains very difficult to study; the return of
processed nuclear matter to the ISM.
This material 
leads to the chemical evolution of galaxies.
Old PNe in decay are the very last objects that can be observed
before the nebular material is fully dispersed and mixes with the 
ISM.\\
Actually this is one of the very few methods to study the properties of the 
ISM directly, in this sense PNe can act as a probe performing an 
active experiment on the ISM.\\

\section{The observations}

\noindent 
Earlier work on interacting PNe by Tweedy \& Kwitter (1996) and Xilouris 
et al. (1996) has given us images of a sample of very large 
(5 to 20 arcmin in diameter) PNe at an angular resolution of about 5 arcsec.
In our survey (Kerber 1998) we have 
concentrated on high resolution imaging
of smaller ($<$ 5 arcmin) PNe combined with long slit spectroscopy.

\begin{table}[h]
\label{tab:oblist}
\centering
\begin{tabular}[t]{lccc} \hline
{Name} & { Identification} & {$\alpha$ (2000.0)} & 
{$\delta$ (2000.0)}\\ 
\hline 
\noalign{\smallskip}
{KLW 11} & { PN G193.0\,$-$\,04.5} & { $05^{h}$ 
$57^{m}$ $08^{s}$} & { $+15^{\circ}$ 25$^{\prime}$ 
31$^{\prime\prime}$}\\ 
{KLW 12} & { PN G197.0\,$+$\,05.8} & { $06^{h}$ 
$43^{m}$ $26^{s}$} & { $+16^{\circ}$ 48$^{\prime}$ 
53$^{\prime\prime}$}\\ 
{EGB 9} & { PN G209.4\,$+$\,09.4} & { $07^{h}$ 
$19^{m}$ $01^{s}$} & { $+07^{\circ}$ 23$^{\prime}$ 
17$^{\prime\prime}$}\\ 
{KeWe 3} & { PN G238.4\,$-$\,01.8} & { $07^{h}$ $33^{m}$ 
$25^{s}$} & { $-23^{\circ}$ 25$^{\prime}$ 44$^{\prime\prime}$}\\ 
{MeWe 1-1} & { PN G272.4\,$-$\,05.9} & { $08^{h}$ $53^{m}$ 
$37^{s}$} & { $-54^{\circ}$ 05$^{\prime}$ 08$^{\prime\prime}$}\\ 
{NeVe 3-1} & { PN G275.9\,$-$\,01.0} & { $09^{h}$ $34^{m}$ 
$01^{s}$} & { $-53^{\circ}$ 11$^{\prime}$ 59$^{\prime\prime}$}\\ 
{Lo 4} & { PN G274.3\,$+$\,09.1} & { $10^{h}$ 
$05^{m}$ $46^{s}$} & { $-44^{\circ}$ 21$^{\prime}$ 33$^{\prime\prime}$}\\ 
{MeWe 1-2} & { PN G283.4\,$-$\,01.4} & { $10^{h}$ $14^{m}$
$24^{s}$} & { $-58^{\circ}$ 11$^{\prime}$ 49$^{\prime\prime}$}\\ 
{NeVe 3-6} & { PN G292.5\,$+$\,00.9} & { $11^{h}$ 
$25^{m}$ $43^{s}$} & { $-60^{\circ}$ 14$^{\prime}$ 
30$^{\prime\prime}$}\\ 
{KFR 1} & { PN G296.3\,$+$\,03.1} & { $12^{h}$
$00^{m}$ $14^{s}$} & { $-59^{\circ}$ 04$^{\prime}$ 
34$^{\prime\prime}$}\\ 
{HaTr 1} & { PN G299.4\,$-$\,04.1} & { $12^{h}$ $16^{m}$ 
$33^{s}$} & { $-66^{\circ}$ 45$^{\prime}$ 46$^{\prime\prime}$}\\ 
{WeKG 2} & { PN G308.4\,$+$\,00.4} & { $13^{h}$ 
$38^{m}$ $42^{s}$} & { $-61^{\circ}$ 55$^{\prime}$ 
45$^{\prime\prime}$}\\ 
{SuWt 1} & { PN G309.2\,$+$\,01.3} & { $13^{h}$ 
$43^{m}$ $59^{s}$} & { $-60^{\circ}$ 49$^{\prime}$ 42$^{\prime\prime}$}\\ 
{MeWe 2-4} & { PN G314.0\,$+$\,10.6} & { $14^{h}$
$01^{m}$ $15^{s}$} & { $-50^{\circ}$ 40$^{\prime}$ 12$^{\prime\prime}$}\\ 
{MeWe 1-4} & { PN G315.9\,$+$\,08.2} & { $14^{h}$
$17^{m}$ $30^{s}$} & { $-52^{\circ}$ 26$^{\prime}$ 19$^{\prime\prime}$}\\ 
{LoTr 10} & { PN G316.3\,$-$\,01.3} &  { $14^{h}$ $46^{m}$ 
$20^{s}$} & { $-61^{\circ}$ 13$^{\prime}$ 30$^{\prime\prime}$}\\ 
{Lo 10} & { PN G328.2\,$+$\,01.3} & { $15^{h}$ 
$49^{m}$ $29^{s}$} & { $-52^{\circ}$ 30$^{\prime}$ 17$^{\prime\prime}$}\\ 
{K 1-31} & { PN G335.4\,$+$\,09.2} & { $15^{h}$ 
$53^{m}$ $13^{s}$} & { $-41^{\circ}$ 50$^{\prime}$ 
25$^{\prime\prime}$}\\ 
{HaTr 3} & { PN G333.4\,$-$\,04.0} & { $16^{h}$ $39^{m}$ 
$38^{s}$} & { $-52^{\circ}$ 49$^{\prime}$ 11$^{\prime\prime}$}\\ 
{KeWe 5} & { PN G348.9\,$+$\,04.6} & { $16^{h}$ 
$57^{m}$ $56^{s}$} & { $-35^{\circ}$ 24$^{\prime}$ 
56$^{\prime\prime}$}\\ 
{Tc\,1}  & { PN G345.2\,$-$\,08.8} & { $17^{h}$ $45^{m}$ 
$35^{s}$} & { $-46^{\circ}$ 05$^{\prime}$ 23$^{\prime\prime}$}\\ \hline
\end{tabular}
\caption{List of low surface planetary nebulae studied in this work}
\label{oblist}
\end{table}

We have collected the largest, homogeneous data set on 
old PNe interacting with the ISM for study of the physical properties
of these objects. In Table\,1 our sample is summarized.
 A more detailed description of the observations and
some individual objects can be found in Rauch et al. (2000 this volume).\\

\section{Observational Results}

In our sample of 21 low surface brightness PNe we have found signs of
interaction with the ISM in 15 cases. 
This unexpectedly large percentage may be the result of an 
observational bias: the interaction leads to a -- usually asymmetric -- 
brightness
enhancement in these low surface brightness objects facilitating their 
discovery, or to put it differently: some of the nebulae may not have been 
discovered if it had not been for the interaction with the ISM.\newline
For at least five of these 15 PNe the central 
star has been found to be displaced from the geometrical center indicating a 
very advanced stage of interaction.
Combined with the fact that the spectroscopically derived electron densities
are low in all cases and very low
for most of the objects (Tab.\,3), this is clear evidence that the interaction 
is common in evolved PNe.
Using long-slit spectroscopy we have for the first time been able to
characterize the plasma parameters of these nebulae demonstrating that
the interaction regions usually show an increased electron density
and -- in most cases -- a pronounced enhancement of the low-ionization stages.
The [N\,{\sc ii}]/H$\alpha$ ratios show absolute values of 1 to 4 with one 
example reaching 12 and increase by factors of up to 3.5 compared with the 
inner parts of the nebulae which are not effected by the interaction.
A correspondingly lower excitation class is also observed. \newline
By combining both imaging and spectroscopy, we are therefore 
able 
to diagnose the degree of interaction. \newline
All of the above is consistent with the current theoretical understanding 
of the interaction process as described by Borkowski, Sarazin, \& Soker  
(1990), Soker, Borkowski, \& Sarazin (1991) and Dgani (2000 this volume). 
A schematic description of the interaction process can be found in 
Rauch et al. (2000).

\begin{table}[h]
\centering
\begin{tabular}[t]{lr@{$\times$}lr@{.}lccc} \hline
{Nebula} & \multicolumn{2}{c}{Diameter}
& \multicolumn{2}{c}{Ratio} & 
{Decentral.} & {Asymm.} & {Sign of}\\ 
 & \multicolumn{2}{c}{$^{\prime\prime}$} & 
\multicolumn{2}{c}{of axes} 
& {central star} & {shape} & {instability}\\ \hline 
\noalign{\smallskip}
{EGB 9} & 377 & 234 & ~~0 & 62 & ? & ++ & ++ \\ 
{KeWe 3} & 283 & 306 & 0 & 92 & ? & ++ & ++\\ 
{MeWe 1-1} & 143 & 165 & 0 & 87 & + & ++ & + \\ 
{NeVe 3-1} & 62 & 53 & 0 & 85 & ? & ++ & ++\\ 
{MeWe 1-2} & 272 & 256 & 0 & 94 & + & + & +\\ 
{KFR 1} & 90 & 100 & 0 & 9 & + & ++ & ++\\ 
{HaTr 1} & 76 & 73 & 0 & 96 & ? & ++ & $-$\\ 
{SuWt 1} & 94 & 53 & 0 & 56 & ? & ++ & $-$ \\ 
{MeWe 2-4} & \multicolumn{2}{l}{~$\sim$ 480} & \multicolumn{2}{c}{$\sim$ 1} & ++ & ++ & +\\ 
{MeWe 1-4} & 113 & 148 & 0 & 76 & ++ & ++ & ++\\ 
{Tc 1} & \multicolumn{2}{l}{~~~~57}  & \multicolumn{2}{c}{1} & + & $-$ & ++\\ \hline
\end{tabular}
\caption{Morphological features in some interacting PNe}
\label{comp_param}
\end{table}

\begin{table}[h]
\centering
\begin{tabular}[t]{lr@{ }lr@{ }lccr@{.}lr@{.}l} \hline
\noalign{\smallskip}
{Nebula} & \multicolumn{4}{c}{$n_{\rm e}$ /cm$^{-3}$} & 
 \multicolumn{2}{c}{Excitation class} & \multicolumn{4}{c}{{$\frac{[N\,II]\;6584}{H{\alpha}\;6563}$}} \\ 
\noalign{\smallskip}
\hline
\noalign{\smallskip}
 & \multicolumn{2}{c}{~inner} & \multicolumn{2}{c}{interaction} & {inner} &  
{interaction} & \multicolumn{2}{c}{inner} &  \multicolumn{2}{c}{interaction} \\
 & \multicolumn{2}{c}{~part} & \multicolumn{2}{c}{zone} & {part} &
{zone} & \multicolumn{2}{c}{part} & \multicolumn{2}{c}{zone} \\ \hline 
\noalign{\smallskip}
{EGB 9} & $<$& 100 & & 170 & 5 & 4 & 0 & 3 & 0 & 8 \\ 
{KeWe 3} & $<$&100 & ~~$<$ & 100 & 3 & 3 & 11 & 2 & ~~~~12 & 7 \\ 
{MeWe 1-1} & $<$&100 & & 180 & 5 & 4 & 0 & 8 & 2 & 7 \\ 
{NeVe 3-1} & &120 & & 220 & 2 & 3 & 0 & 8 & 2 &5 \\ 
{MeWe 1-2} & $<$&100 & & 260 & 4 & 2 & 1 & 7 & 3 & 0 \\ 
{KFR 1} & & 250 & $<$ & 100 & 4 & 4 & 1 & 9 & 1 & 6 \\ 
{HaTr 1} & $<$ & 100 & & ~~-- & 5 & -- & 0 & 5 & \multicolumn{2}{c}{--} \\ 
{SuWt 1} & & 170 & & 200 & 7 & 4 & 2 & 1 & 3 & 8 \\ 
{MeWe 2-4} & & ~~-- & & 390 & 6 & 5 & 0 & 8 & 2 & 7 \\ 
{MeWe 1-4} & & 470 & & 120 & 6 & 3 & 0 & 5 & 0 & 8 \\ \hline
\end{tabular}
\caption{
Nebular parameters from plasma diagnostics}
\label{comp_plasma}
\end{table}

Morphologically our deep high-resolution images show an enormous amount of
detail and clearly show indication of instability, in some cases the nebulae
are obviously in the process of being broken apart, for 
example see the images of MeWe\,1-4 and KFR\,1 in Rauch et al. (2000).\newline
It has been shown recently by 
Soker \& Dgani (1997) and Dgani \& Soker (1998)
that the effect of the interstellar medium's magnetic field (ISMF) can be 
extremely important. 
We see evidence for the ISMF's action in some of our objects
in the form of stripes or rolls as described by Dgani (2000)
in this volume.

\section{Future work}
In this project we have already begun to extend the sample to
the northern hemisphere and have included very large objects 
requiring wide-field imaging. Another important aspect will be the study of
the central stars giving us information on their evolutionary status, as well
as the spectroscopic distance.

\section*{Acknowledgments}
This work was supported by travel grants from the Austrian Ministerium f\"ur
Wissenschaft, Forschung und Verkehr and the Universit\"at Innsbruck and
by the DLR under grant 50 OR 9705 5 (TR).


\begin{references}
\reference Borkowski K.J., Sarazin C.L., \& Soker N., 1990, ApJ 360, 173


\reference Dgani R., \& Soker N., 1998, ApJ 495, 337

\reference Kerber F., 1998, Rev. in Mod. Astronomy 11, ed. 
R.E. Schielicke, Astronomische Gesellschaft, Jena, p.\,161


\reference Kwok S., Purton C.R., \& Fitzgerald P.M., 1978, ApJ 219, L\,125

\reference Soker N., \& Dgani R., 1997, ApJ 484, 277 

\reference Soker N., Borkowski K.J., \& Sarazin C.L., 1991, AJ 102, 1381

\reference Tweedy R.W., \& Kwitter K.B., 1996, ApJS 107, 255

\reference Xilouris K.M., Papamastorakis J., Paleologou E., \& Terzian Y.,
1996, A\&A 310, 603

\end{references}
\end{document}